\documentstyle{article}
\begin{document}
\setlength{\baselineskip}{14pt}
\def\theequation{\thesection.\arabic{equation}}
\newcommand{\rf}[1]{(\ref{#1})}
%
%  Make Title Page
%
\setcounter{page}{1}
\null
\vskip 5pt
\hfill UCSBTH-93-03
\vskip 5pt
\begin{center}
   {\LARGE
Complex Numbers, Quantum Mechanics and the Beginning of Time
    \par}
    \vskip 3em
    {\large
     \lineskip .75em
     \begin{tabular}[t]{c}
G. W. GIBBONS
\\
\\
D.A.M.T.P. \\
University of Cambridge \\
Silver Street  \\
Cambridge CB3 9EW \\
U.K.
\\
\\
\\
H.-J. POHLE
\\
\\
Department of Physics\\
University of California\\
Santa Barbara, Ca 93106
\\
\\
January 12,1993
\\
    \end{tabular}\par}
    \vskip 4em
\end{center}
\begin{abstract}
A basic problem in quantizing a field in curved space is the decomposition
of the classical modes in positive and negative frequency. The
decomposition is equivalent to a choice of a complex structure in the
space of classical solutions. In our construction the real tunneling
geometries provide the link between the this complex structure and
analytic properties of the classical solutions in a Riemannian section of
space. This is related to the Osterwalder- Schrader approach to Euclidean
field theory.
\end{abstract}
\newpage
\section{Introduction}

In quantum gravity are seeks to unite in  a  harmonious  whole  two  distinct
theories, classical  general  relativity  and  quantum  mechanics.  It  seems
reasonable to suppose that what eventually emerges will entail a  substantial
revision of  both  theories  so  that  both  quantum  mechanics  and  general
relativity emerge merely as approximations to some underlying  exact  theory.
In the context of cosmology for example the idea of a classical spacetime  in
which all conventional laws of quantum mechanics hold would expected to be  a
good approximation at large times.

In fact classical relativity already brings about a radical  modification  of
the idea of  time  used  in  non-relativistic  physics  while  compared  with
classical theory one of the striking new features of quantum mechanics is its
use of complex structures, particularly in the description of time-evolution.
Thus it seems reasonable to expect that it is in the way that complex
structures enter quantum mechanics where modifications due to quantum gravity
will arise. One possible viewpoint is that the complex structure of the
quantum mechanical Hilbert space only emerges late in the history of the
universe.

Of course in the absence of a complete theory of quantum gravity one can only
speculate about such things unless one tries to construct approximate  models
in which the well  known  divergence  difficulties  of  quantum  gravity  are
ignored in the hope that in the more fundamental theory the divergencies  are
eliminated while at the same time the qualitative features of  the  model  at
large scales persist. If one cannot construct approximate  models  exhibiting
certain features at large times, it seems unlikely that such  features  could
arise in any more fundamental theories. If one can, strictly speaking one can
say nothing except perhaps to confess to a feeling of optimism. This seems to
be the prevailing view in quantum cosmology and in that spirit we  are  going
to investigate a simple-minded model for the quantum creation of the universe
mediated by real tunneling geometries~\cite{hartle-gibbons}.

These arise in the WKB ansatz for the solution of the Wheeler-DeWitt equation
as a special case of the general situation in which  one  considers  "complex
paths" which are spacetimes with complex metrics. In the WKB ansatz one  uses
a fixed background geometry and questions  about  the  complex  structure  of
quantum mechanics are equivalent to asking how the complex structure  of  the
quantum mechanical Hilbert space for quantum fields around a fixed  spacetime
is determined, quantum fields correspond  to  the  fluctuations  around  that
background.

The decomposition into positive- and negative-frequency parts is
related to a natural complex structure $J$. It is $J$ that determines what we
mean by complex numbers in the quantum theory of a classical system.  For
Lorentzian spacetime this is a problem studied by Ashtekar and Magnon
\cite{ashtekar} and we have been greatly influenced by that paper.

Real tunneling geometries are partially Lorentzian and partially  Riemannian.
Alternatively the Lorentzian  and  Riemannian  portion  may  be  regarded  as
different real slices of a complex spacetime $M_C$, which have a common
boundary $ \Sigma $. This surface $\Sigma$ acts as an initial Cauchy surface
for the Lorentzian spacetime and its quantum fields ($\Sigma$ might be called
the "beginning of time").

Real tunneling geometries give a privileged notation of positive and
negative frequency and hence a privileged "vacuum" state brought about the
tunneling process.  The choice of positive frequency or of a complex
structure may be thought of as a direct sum decomposition of the boundary
data on $\Sigma$ into data which evolve to give a solution everywhere
bounded in $M_{R}^{-}$ or everywhere bounded in $M_{R}^{+}$, where
$M^{+}_{R}$ and $M^{-}_{R}$ are the two halves of the Riemannian slice
separated by $\Sigma$.  In this way our construction associates in a very
clear way the complex structure of quantum mechanics with the beginning of
time.

Additionally using the ideas of Euclidean Quantum Field Theory generalized to
this particular class of curved spacetime the complex structure of the
quantum mechanical Hilbert space is intimately related to Osterwalder and
Schrader's use of Reflection Positivity to construct the Quantum Mechanical
Hilbert Space  \cite{osterwalder-schrader}, \cite{uhlmann},
\cite{gibbons-schmutzer}, \cite{jaffe}
and \cite{angelis}. In that construction one considers reflections about a
spacelike hypersurface. This surface corresponds in our construction to the
boundary $ \Sigma $.

We shall show in section~\ref{sec-example} that this agrees with previous
work in the simplest possible case - that of the deSitter spacetime. We also
apply it to the Page metric which is also a Real Tunneling Geometry but one
which is not as symmetric as deSitter spacetime.

In section~\ref{sec-example} we shall also indicate how the formalism we have
developed in section~\ref{sec-tunnel} may be applied to other tunneling
geometries.
In all these sections we will restrict ourselves to Klein-Gordon fields.
In section~\ref{sec-spinor} we treat spinors.
section~\ref{sec-conclusion} is a conclusion.

\section{Complex Structures}
\label{sec-complex}
\setcounter{equation}{0}

In order to clarify the purpose of the work described below we shall review
the relation between time and the complex numbers in quantum mechanics from
the particular perspective we have adopted in this paper. Our basic viewpoint
is that classically we start with real variables - for instance real valued
classical fields - while quantum mechanically we deal {\sl in an essential
way } with complex variables. Of course we often use complex numbers in
classical mechanics but this is merely as a book-keeping device and has no
fundamental significance.

In conventional quantum mechanics it is a basic postulate that
physical states correspond to rays in a complex Hilbert space. One way of
seeing why this must be so in the standard formulation is that "observables"
have a dual role. On the one hand they give the outcomes of measurements in
physical states and thus correspond to quadratic forms. On the other hand
they generate infinitesimal transformations in the space of physical states
into themselves and thus correspond to linear maps of endomorphisms.  If the
space of physical state vectors is viewed as a real vector space $V$
whose $\Phi$ vectors have real components
$
   \phi^a, a = 1,2, \ldots, m = \dim_R(V)
$
then the observables should be regarded as second rank symmetric covariant
tensors with components:
\begin{equation}
   O_{ab}   =  O_{ba},
\label{eq:oab}
\end{equation}
in their first role quadratic forms, and once contravariant once covariant
tensors $T^a \, _b$ in their second role. The vector space ${V}$
possesses a distinguished positive definite observable, the
quantum-mechanical metric
\begin{equation}
   g_{ab}=g_{ba}.
\label{eq:gab}
\end{equation}
Regarding positive semi-definite observables as mixed states or density
matrices one may view the metric $ g_{ab}$ as the density matrix associated
to complete ignorance. In any event the expectation value of the observable
$O_{ab}$ in the pure state with components $\phi ^a$ is
\begin{equation}
   <O>   =  \phi^a O_{ab} \phi^b /
            \phi^a g_{ab} \phi ^b.
\label{eq:expo}
\end{equation}
Clearly state vectors which differ by a real multiple give rise to the same
expectation values for all observables and hence the physical states would
seem on the face of it to correspond to the real projective space
$P({V})  \equiv   P_{m}(R)$.
Clearly every observable in the first sense can be diagonalized with respect
to the positive definite metric $g_{ab}$ over the reals (using orthogonal
transformations) and thus has real eigen-values.

In their second role as infinitesimal transformations observables generate
rotations of ${V}$ into itself which preserve the metric $g_{ab}$. Thus
they satisfy:
\begin{equation}
   T_{ab}   \equiv   g_{ac} T^c\,_b    =  -T_{ba}
\label{eq:tab}
\end{equation}

In fact every linear bijection of the convex cone of positive semi-definite
observables preserving the metric $g_{ab}$ and the trace $ g^{ab} O_{ab}$ is
induced by an orthogonal transformation. Thus the assumption that physical
transformations are orthogonal transformations seems to be well founded,
though one might wish to relax the condition the map of density matrices to
density matrices be bijective.

How can one link the antisymmetric tensor $T_{ab}$ to the symmetric tensor
$O_{ab}$? One way (and this is the way it is done in standard quantum
mechanics) is to invoke the existence on $V$ of a symplectic form,
i.e. a fundamental anti-symmetric tensor:
\begin{equation}
   \Omega _{ab}   =  - \Omega _{ba}
\label{eq:omega}
\end{equation}
which is non-degenerate in the sense that it provides an isomorphism between
$V$ and its dual space $V^*$.  This imposes the constraint on
the real dimension $m$ of the vector space $V$ that it be even. If we
demand that in addition to preserving the metric $g_{ab}$ physical
transformations preserve the symplectic form it follows that
\begin{equation}
   O[{T^b _c}] _{ac}    \equiv   \Omega _{ab} T^b \,_c   =   O_{ca}
\label{eq:otbc}
\end{equation}
Since the metric $g_{ab}$ is really just another observable it should be the
case that there is an infinitesimal transformation associated to it. Let us
call this $J^a \, _b $.  It is defined by:
\begin{equation}
   g_{ac}   =  \Omega _{ab}   J^b \, _c
\label{eq:gac}
\end{equation}
In standard quantum mechanics the tensor $J^a \, _b$ coincides with what is
usually known as a complex structure. However in order that this be so
$J^a \, _c$
must satisfy the condition that:
\begin{equation}
   J^a \, _c   J^c \, _b   =  - \delta ^a _ b
\label{eq:jac}
\end{equation}
where $ \delta ^a _ b$ is the Kronecker delta. This condition imposes a
compatibility condition on the metric $g_{ab}$ and the symplectic form
$\Omega_{ab}$ which may be expressed in various ways. For example $J^a \ _b$
is not only an infinitesimal isometry of the metric $g_{ab}$ but a finite one
as well:
\begin{equation}
   g_{ab}   J^a \, _c   J^b \, _ d  =   g_{cd}.
\label{eq:gabj}
\end{equation}
Similarly it is not only an infinitesimal symplectic transformation but a
finite one as well:
\begin{equation}
   \Omega_ {ab}   J^a \ _c    J^ b \ _ d  =  \Omega _{cd}.
\label{eq:omegaab}
\end{equation}
{}From a physical point of view the simplest justification of the compatibility
conditions would seem to be one of economy. If it were not true then
successive powers of the endomorphism $J^a \, _b $ would, when contracted with
the metric or the symplectic form, produce a further fundamental symmetric,
or antisymmetric tensors with no obvious physical interpretation. In any
event given that the compatibility conditions hold it now follows that
observables must be hermitean, in the sense that:
\begin{equation}
   O_{ab}   J^a \ _c    J^b \ _d    =  O_{cd}
\label{eq:oabj}
\end{equation}

It is at this stage that the quantum mechanical phases as opposed to $\pm 1$
factors enters the formalism. One now has that the state vector $\phi ^a$ and
the state vector
$ \exp ( \alpha J )  ^a \, _b    \phi ^b  $
give the same expectation values for all (hermitean ) observables $O_{ab}$
and real phases $\alpha$.  Thus if one is interested in the question of how
these phases enter into quantum mechanics it is precisely when we introduce
the complex structure.

It is important to realize that given a metric $g_{ab}$ or a symplectic form
$\Omega _{ab}$ there is no unique compatible complex structure associated to
it. In fact it is not uncommon to impose more than one complex structure on
the same real vector space. For example one may start with the 4-dimensional
space of real Majorana spinors ( using a signature for the metric in which
$  \gamma_0 ^2    = -1, $
and
$  \gamma _5 =    \gamma _0   \gamma _1   \gamma _2   \gamma_3 $)
and regard it as a 2-dimensional complex vector space by choosing $\gamma _5$
as the complex structure, in which case one arrives at Weyl spinors, or one
may choose $\gamma _0$ as the complex structure in which case one arrives at
(non-relativistic) Pauli spinors.

One way in which complex structures arise naturally is when one has a one
parameter subgroup $R(t)$ of the orthogonal group $SO(m)$ acting on the real
vector space $V$. For a generic rotation $R(t)$ there will be $m/2$
mutually orthogonal real 2-planes such that the rotation is through an angle
$\lambda_i t$ in the i`th 2-plane. The quantities $\lambda _i$ are the skew
eigen-values of the infinitesimal generator $T^a \, _b$ of the rotation
$  R(t)  =  \exp (   t  T^a \, _b   )  $.
The rotation $R(t)$ commutes with any one of the $2^{m \over 2}$ possible
complex structures defined by rotations of $\pm {\pi \over 2}$ in these
orthogonal 2-planes.  With respect to any of these complex structures the
rotation $R(t)$ is a $U(1)$ subgroup of $U({m \over 2})$. There is an
associated conjugation operator with each such choice. Now if the rotation
$R(t)$ where thought of as time translations, the eigen-values $\lambda$
being thought of as energies, then the complex structure with all positive
signs is precisely the "$i$" of quantum mechanics and the associated
conjugation operator is the time reversal operator. In this way one may say
that a particular quantum mechanical Hamiltonian picks out or fixes the
complex structure. Note that by an appropriate choice of basis we could
always choose the complex structure so as to arrange that the energy
eigen-values were all positive.  However if the complex structure is given
ahead of time then of course the signs of the energy eigenvalues are
determined.

Let us now consider the situation in which there is an additional one
parameter subgroup of rotations $R ^ \prime (t)$, for example that
associated with some conserved charge. In the standard formalism as we have
described it above $R^ \prime (t)$ must commute with the "$i$" of quantum
mechanics, i.e. it must be a $U(1)$ subgroup of $SO(m)$ with respect to the
unitary structure determined by $J$. Now $R ^ \prime (t)$ determines its own
complex structure, call it $J ^ \prime$, and its own conjugation operator,
call it $C$.  The operators $J$ and $J ^\prime$ commute . The operators $-$
and $J$ anti-commute. Similarly the operators $J^ \prime $ and $C$
anti-commute.  However $C$ {\sl commutes } with $J$, that is the charge
conjugation operator $C$ is a linear operator with respect to the usual
complex structure of quantum mechanics, unlike the time reversal operator. We
could of course introduce a complex notation associated to the complex
structure $J ^ \prime$ but this would be confusing because we would have
between the "$i$" of quantum mechanics and this new $i^\prime$, they are
after all not the same operator. In classical field theory where the "$i$" of
quantum mechanics does not enter one frequently does use a complex notation
with the usual "$i$", when dealing with gauge transformations for example,
and no confusion arises but in quantum mechanics this is not possible.

We restrict ourselves to Klein-Gordon fields, classically defined by the
equation

\begin{equation}
        (-\nabla^2_{g} + m^2)\, \Phi =0
\label{eq:klein_gordon}
\end{equation}
on a manifold $M$ with Lorentz metric $g$. Let $V$ denote the vector-space of
all
well-behaved real-valued solutions of equation \rf{eq:klein_gordon}. On $V$
we can define a antisymmetric bilinear form $\Omega$ by means of the integral

\begin{equation}
   \Omega(     \Phi_1,  \Phi_2   ) =
   \int_\sigma
   ( \Phi_2 \nabla_\mu \Phi_1 - \Phi_1 \nabla_\mu \Phi_2 )
   d \Sigma^\mu
\label{eq:omega-integral}
\end{equation}
over a surface $ \sigma $, where $\Phi_1$ and $\Phi_2$ are in $V$.

The one-particle Hilbert space $\cal H$ is a copy of $V$. But to represent
quantum states of a particle, $\cal H$ must have, in addition, the structure
of a complex Hilbert space. So we have to introduce on $V$ a hermitean
inner product together with a complex structure $J$, which is compatible with
the symplectic structure $\Omega$, i.e. $ J^2 = -1$ and

\begin{equation}
   \Omega(  J \Phi_1,J \Phi_2 )  =\Omega(    \Phi_1,  \Phi_2   )  .
\label{eq:compatible}
\end{equation}
If \( \Omega ( ., J .) \) is non-degenerate then it automatically follows
that
\begin{equation}
   \langle  \Phi_1, \Phi_2    \rangle  =
   \Omega ( \Phi_1, J \Phi_2  ) + i\Omega ( \Phi_1, \Phi_2 )
\label{eq:scalar-product}
\end{equation}
provides a Hilbertian norm on $V$.

Given any complex structure $J$, regardless whether or not it is compatible
with a symplectic form, one may introduce the standard complex notation in
which the
operator $J$ becomes multiplication by "$i$". Formally this proceeds
as follows. One first complexifies the real vector space ${V}$ by tensoring
with an algebra which is isomorphic to the complex numbers and which we
denote for the time being by $C$ with imaginary unit denoted as usual by $i$
to obtain a complex vector space,
\begin{equation}
   V_C   =  V  \times_R    C.
\label{eq:vc}
\end{equation}
One extends the action of $J$ to $ V $ in a $C-$linear fashion. One
then has a preferred direct sum decomposition of $V_C$ :
\begin{equation}
   V_C   =  {\cal H}    \oplus   {\overline {\cal H} } ,
\label{eq:preferred-decomposition}
\end{equation}
where
\begin{equation}
   {\cal H} =  {1 \over 2} (1-iJ)   V_C
\label{eq:h12}
\end{equation}
and
\begin{equation}
   {\overline {\cal H}}    =  { 1 \over 2}   (1+iJ)   V_C.
\label{eq:h12bar}
\end{equation}
It is easy to see that the action of $J$ on any element of $\cal H$ is just
given by multiplication with $i$.  Associated to the complex structure $J$ is
a complex conjugation operator (written as $-$ ) which anti-commutes with the
operator $J$. Any real vector of $V$ can now be decomposed as \mbox{\( \Phi =
\Phi^+ + \Phi^- \) }. The vector $\Phi^-$ is the complex conjugate of
$\Phi^+$. Although $\Phi$ and $J \Phi$ are real, $\Phi^+$ and $\Phi^-$ are
complex.  Thus physically $\cal H$ has as many "degrees of freedom" as the
original vector space.

The quantum mechanical Hilbert space is then
given as the space of analytic functions, where analyticity is defined with
respect to $J$ and the scalar product  in $\cal H$ is

\begin{equation}
         \langle  \Phi_1, \Phi_2    \rangle  =
         i \, \Omega(\Phi_1 , \overline{\Phi_2} ) .
\label{eq:scalar-hilbert}
\end{equation}
Of course it has to be shown in every special case, that the product
\rf{eq:scalar-hilbert} is positive definite.

In a static spacetime, with $\sigma$ in equation \rf{eq:omega-integral} to be
a Cauchy hypersurface orthogonal to the time-like Killing field, the
procedure above decomposes every real solution $\Phi $ of the Klein-Gordon
equation into positive and negative frequency parts
\mbox {  \( \Phi = \Phi^+ + \Phi^-  \) }.
This decomposition provides a preferred complex structure which is given by
\mbox {  \( J\Phi = i \Phi^+  +  (-i)  \Phi^-\) }.

Finally, to obtain a description of quantum fields in curved spacetime we
have to construct the Hilbert space of states $ \cal F $, as a sum of
symmetrized tensor products:

\begin{equation}
{\cal F} = C \oplus {\cal H} \oplus (\,{\cal H} \otimes_S
           {\cal H} ) \oplus ...
\label{eq:hproduct}
\end{equation}
The summand $C$ in \rf{eq:hproduct} is a vacuum or ground state.

On the level of the bosonic Fock space we have to associate with each pair
of solutions
\mbox{ $\Phi^+_k, \Phi^-_k $ } of $\cal H$ or $\overline{\cal H}$ resp.
creation and destruction operators
\mbox{ ${ \bf a}_k,  { \bf a}^\dagger_k  $}
and to impose the canonical quantization relation
\begin{equation}
   [\,   {\bf a}_k,  {\bf a}^\dagger_{k'} ]  =  \delta_{kk'}
\label{eq:vertauschung-allgemein1}
\end{equation}
and
\begin{equation}
   [\,   {\bf a}_k,{\bf a}_{k'} ]  =  0  \ \ \ \ \ \
   [\,   {\bf a}^\dagger_{k},  {\bf a}^\dagger_{k'}  ]  =  0.
\label{eq:vertauschung-allgemein2}
\end{equation}

Now we can associate with each element $\Phi$ of $V$ a field-operator
$  {\bf \Phi}$ by means of the relation
\begin{equation}
   {\bf \Phi}   =  \sum_k
         (  {\bf a}_k \Phi_k^- + {\bf a}^\dagger_k \Phi_k^+ ),
\label{eq:field-operator-allg}
\end{equation}
where the modes are normalized with respect to \rf{eq:scalar-hilbert}.
  The vacuum state is defined by
% %
\begin{equation}
   {\bf a}_k | 0 \!>   = 0.
\label{eq:vacuum}
\end{equation}
Now  we see that the definition of a complex structure $J$ is crucial
for the definition of the field-operator and the vacuum state..
The Wightman function is given by the vacuum expectation-value of two
field-operators:
\begin{equation}
   W(x,y)   =  <0|{\bf \Phi} (x){\bf \Phi} (x)|0 \!>  =
      \sum_k \Phi_{k}^-(x) \Phi_{k}^+(y)
\label{eq:wightman}
\end{equation}

The basic example we have in mind is that of a  simple  harmonic  oscillator.
Quantum Field Theory, from that point of view, is just a collection  of  many
such oscillators.

The oscillator position $y$ satisfies:
\begin{equation}
   \frac{d^2 y}{dt^2}   +  \nu^2 y  = 0
\label{eq:harmon}
\end{equation}
with $\nu$ real and positive. A basis for the real 2-dimensional space
$V$ of classical solutions is provided by
\begin{equation}
   e_1   = \frac{1}{\sqrt \nu}   \cos \nu t, \ \ \ \ \ \
   e_2   = \frac{1}{\sqrt \nu}   \sin \nu t.
\label{eq:basis}
\end{equation}
Their Wronskian is independent of time
\begin{equation}
   \Omega   (e_1,e_2)   = e_1 \dot{e_2}   -  e_2 \dot{e_1}  =  1
\label{eq:wronski}
\end{equation}
and gives the symplectic form on the real 2-dimensional vector space
$V$.
A complex structure on $\cal{H}$ is induced from the complex structure $J$ on
the space of real solutions $V$ given by:

\begin{equation}
   J e_1=e_2 \ \ \ \ \  Je_2=-e_1
\label{eq:complex_structure}
\end{equation}
so that

\begin{equation}
   J \left( \frac{1}{\sqrt \nu}  \, e^{\mp i \nu t} \right) =
   (\pm i)  \frac{1}{\sqrt \nu}  \, e^{\mp i \nu t}.
\label{eq:J_auf_e-}
\end{equation}
The norm of a quantum mechanical state
\mbox{ \( \Phi=\phi_1 e_1+\phi_2 e_2   \)}
is \mbox{ \(   \| \Phi \|  = \phi_1^2 + \phi_2^2   \)}.

The usual procedure in second quantization (in the Heisenberg picture) is to
regard $y$ as an operator $\bf y$

\begin{equation}
   {\bf y}  =  {\bf a}        \frac{1}{\sqrt \nu}  \, e^{-i \nu t}   +
               {\bf a}^\dagger \frac{1}{\sqrt \nu} \, e^{i \nu t}
\label{eq:yhut}
\end{equation}
and impose the canonical quantization relation
\begin{equation}
   [\,   \bf{a},  \bf{a}^\dagger ]  =  1
\label{eq:vertauschung}
\end{equation}
In the Bargmann-Fock representation we  consider  wave  functions  which  are
anti-holomorphic functions of $a$, $\Phi(\overline{a})$. One has
\begin{equation}
   \bf{a}^\dagger \rightarrow \overline{a}, \ \ \ \ \
   \bf{a}         \rightarrow    \frac{\partial}{\partial\overline{a}}.
\label{eq:trafo}
\end{equation}
The inner product is
\begin{equation}
   <  \Phi_1   | \Phi_2 >  \: =  \int da \,d \overline{a}   \frac{1}{2 \pi i}
   e^{-a \overline{a}   }  \,
   \overline{ \Phi_1( \overline{a} )}  \Phi_2( \overline{a}).
\label{eq:aproduct}
\end{equation}
We may identify $\cal H$ with the space of anti-holomorphic functions of $a$.
The ground state correspond to the constant function.  The first excited
state, corresponding to $\cal H $ is given by $ \Phi_1 = \overline{a} $.
The n'th exited state has wave functions $ \Phi_n = \overline{a}^n /
\sqrt{n!}   $
and corresponds to
$  (  \cal{H} \otimes   \ldots   \otimes \cal{H}   )_S   $,
the n-fold symmetric product. The action of $J$ on $\Phi$ is
\begin{equation}
   J \Phi(\overline{a}) = i   \Phi(\overline{a}).
\label{eq:J_auf_psi}
\end{equation}

\section{Globally Static Metrics}
\setcounter{equation}{0}
\label{sec-static}

In this section we reanalyse the complex structure in a globally static
spacetime and its connection with analytic properties of the analytic
continuation of positive and negative frequency modes in a corresponding
Riemannian space.

In a globally static Lorentzian spacetime $M_L$ for which the metric may be
written as:

\begin{equation}
 ds^{2} = - v ( {\bf x} ) dt^{2}
                 + g_{ij}({\bf x} ) \,dx^{i} dx^{j}
 \;,\;\;v>0\;,
\label{eq:stat_metric}
\end{equation}
a basis for the space $V$ of real solutions of the Klein- Gordon equation is
given by
\begin{equation}
   \Phi_c({\bf x},t;\omega)   =
   \cos (\omega t)\, \chi_\omega \!( {\bf x} )
   \;,\;\; \omega>0
\label{eq:cos_stat}
\end{equation}
and

\begin{equation}
   \Phi_s({\bf x},t;\omega)   =
   \sin (\omega t)\, \chi_\omega \!( {\bf x} )
   \;,\;\; \omega>0
\label{eq:sin_stat}
\end{equation}
where
$\chi_\omega \!( {\bf x} )$
satisfies the spatial equation:

\begin{equation}
 -\frac{1}{\sqrt{v}} \nabla_{i} (\sqrt{v} \nabla^{i} \chi_\omega )
 +\frac{\omega^{2}}{v} \chi_\omega
 +m^{2} \chi_\omega = 0 .
\label{eq:sep_spatial_equ}
\end{equation}

We can endow the space $V$ of real valued solutions of the Klein- Gordon
equation with the symplectic form

\begin{equation}
\Omega (\Phi,\Phi') = \int_{\Sigma} (  \Phi \partial_{\mu} \Phi'
                              - \Phi' \partial_{\mu} \Phi ) d \Sigma^{\mu}
\label{eq:spl}
\end{equation}
the integral being taken over a Cauchy surface $\Sigma$. To convert $V$
to the one- particle Hilbert space ${\cal H}$ we introduce the complex
structure $J$ with
\begin{equation}
   J \Phi_c =  \Phi_s \mbox{ \ \ \ \ \ and \ \ \ \ \ \ }
   J \Phi_s = - \Phi_c,
\label{eq:globally-complex}
\end{equation}
and perform the preferred direct sum decomposition
\rf{eq:preferred-decomposition}. We obtain the space $\cal H$ with the basis

\begin{equation}
   \Phi^+ ({\bf x},t; \omega) =
   \exp (- i \omega t)\, \chi_\omega \!( {\bf x} )
   \;,\;\; \omega>0  .
\label{eq:exp_stat}
\end{equation}
This is the "obvious" definition of positive frequency. If we decompose a
solution $\Phi$ of $V$ in its positive and negative frequency part we find
that our complex structure is given by
\begin{equation}
   J\Phi = i \Phi^+  +  (-i)  \Phi^-   .
\label{eq:globally-complex-kurz}
\end{equation}

Setting $t=i \tau$, $\tau$ real, we pass to a metric of a
Riemannian space $M_R$. We regard $M_L$ and $M_R$ as real slices of a complex
manifold $M_C$ with the common boundary $\Sigma$.

We see that a superposition $\Phi$ of the analytic continuation of purely
positive frequency functions satisfies the elliptic Klein-Gordon equation

\begin{equation}
 -\frac{1}{\sqrt{v}} \nabla_{i} (\sqrt{v} \nabla^{i} \Phi )
 -\frac {1}{v} \frac { \partial^{2} \Phi } { \partial \tau^{2} }
 +m^{2} \Phi = 0
\label{eq:spatial_equ}
\end{equation}
for all negative values of $\tau$, i.e. it is everywhere bounded on
\( M_{R}^{-} \), defined by
$\tau<0$. In other words ${\cal H}$ analytically continues to the space
$H^{+}$ of solutions of the homogeneous Klein-Gordon equation bounded in
$M_{R}^{-}$ similarly $\overline{{\cal H}}$, i.e.  the negative frequency
solutions, analytically continue to solutions in $H^{-}$ of the homogeneous
Klein-Gordon equation which are bounded on $M_{R}^{+}$ i.e.  if $\tau>0$.
The reflection map $\theta:\tau\rightarrow-\tau$ maps the
two spaces $H^+$ and $H^-$ into one another:

\begin{equation}
 \theta H^{\pm} = H^{\mp}
\label{eq:theta_h}
\end{equation}

Now we can give a more geometrical motivation for the complex structure
introduced in \rf{eq:globally-complex}:
Instead of referring to the space of solutions, either of the
Lorentzian or the Riemannian Klein-Gordon equation we may  refer  instead  to
their values and the values of their normal derivatives ( either
$\frac{ \partial \Phi}{ \partial t}$ or
$\frac{ \partial \Phi}{ \partial \tau}$
on the boundary surface $\Sigma$, i.e. at $\tau=0$. In the Lorentzian case
this  boundary
data is the (complex-valued) Cauchy data $V_{C}$.  In  the  Riemannian  case
the boundary data provide a real vector space of pairs (
\(  \Phi_{|\Sigma}, \frac{\partial \Phi}{\partial \tau}_{| \Sigma} \)
)
The choice of positive frequency or of a complex structure may thus be
thought of as a direct sum decomposition of the boundary data on $\Sigma$
into data which evolve to give a solution everywhere bounded in $M_{R}^{-}$
(elements of $H^{+}$ ) or everywhere bounded in $M_{R}^{+}$ (elements of
$H^{-}$ ). In this way our construction associates in a very clear way the
complex structure of quantum mechanics with the beginning of time.

Note that while ${\cal H}$ and $\overline{{\cal H}}$ may be thought of as
complex conjugate of one another, $H^{+}$ and $H^{-}$ are both real vector
spaces.  Note also that the decomposition of the real boundary data into \(
H^{+} \oplus H^{-} \) is non-local with respect to the boundary $\Sigma$.
In a quite recent paper Ashtekar, Tate and Uggla \cite{ashtekar-tate} showed,
that complex structures can be constructed from Killing vectors. So the
complex structure for globally static spacetime is very related to the
time-like Killing vector in these geometries.

\section{Real Tunneling Geometries }
\label{sec-tunnel}
\setcounter{equation}{0}

The aim of this section is to extend the theory developed in
section~\ref{sec-static} to the more general case when the Lorentzian
spacetime, which we call $2M_L$, admits a moment of time symmetry $\Sigma$.
Thus $2M_L$ can be decomposed as

\begin{equation}
   2M_L = M_L^+ \cup M_L^-
\label{eq:decompose}
\end{equation}
where time reversal $T$ is an isometry of $g_L$ such
that
\begin{equation}
   T M_L^\pm = M_L^\mp
\label{eq:t_symmetry}
\end{equation}
and
\begin{equation}
   \partial M_L^+ = \Sigma = \partial M_L^-
\label{eq:sym_sigma}
\end{equation}
where
\begin{equation}
   \partial \Sigma=\Sigma
\label{eq:def_sigma}
\end{equation}
We assume (following) that in the compactification $M_C$ of $2M_L$ there is
another real section $2M_R$ with Riemannian metric $g_R$ and a reflection
map $\theta$. The conditions \rf{eq:t_symmetry} - \rf{eq:def_sigma} continue
to hold but with $L$ replaced by $R$ and $T$ replaced by $\theta$. The
real tunneling geometry is the manifold:
\( M_R^- \cup M_L^+  \)
with the metric \(   g_R   \) on $M_R^-$ and $g_L$ on $M_L^+$.
The surface $\Sigma$ thus represents the "beginning of time" in the model.
In \cite{hartle-gibbons} it was further assumed that $M_R^-$ was compact and
hence the
boundary $\Sigma$ had to be compact as well. In our work we shall not
necessarily be making that assumption. However we shall assume that $\Sigma$ is
the only boundary of $M_R^-$. This assumption would rule out the globally
static example discussed in section~\ref{sec-static} for which
\( M_R^-=\Sigma \times (-\infty,0]  \)
, $\Sigma$ being the spatial cross section and \(  (-\infty,0] \)
corresponds to
\( -\infty < \tau \leq 0   \).
Demanding that various fields vanish at
\( \tau \rightarrow - \infty \) we have the effect of rendering irrelevant
the non-compactness of $M_R^-$ and in fact our results will cover that more
general case as well. All that is really needed for that the only boundary
terms we need to consider are those on $\Sigma$.

We now proceed very much as in section~\ref{sec-static}. We define the
spaces $H^+$ and $H^-$ of real solutions of the equation

\begin{equation}
   (-\nabla^2_{g_R} + m^2)\, \Phi_R =0
\label{eq:klein_gordon2}
\end{equation}
which are everywhere bounded on $M_R^-$ or $M_R^+$ respectively. We will
identify these spaces with the boundary data
\( (\Phi_R, \frac{\partial \Phi_R}{\partial \tau})    \)
where the normal derivative
\( \frac{\partial \Phi_R}{\partial \tau}  \)
is outward with respect to $M_R^-$ and inward with respect to $M_R^+$.
The map $\theta$ extends to a map from
\begin{equation}
   \theta : H^\pm \rightarrow H^\mp
\label{eq:theta_h2}
\end{equation}
by pull back, that is for all
\( \Phi_R \in H^ \pm \)
we define
\begin{equation}
   \theta \Phi_R(x) = \Phi_R (\theta x) = \Phi_{R \theta} (x) \;.
\label{eq:theta_f}
\end{equation}
Now we perform an analytic continuation of solutions in $H^+$ on $M_R^-$
to get complex solutions in $M_L^+$. These solutions define a vector space
$\cal H$ and we consider these solutions as to be of "positive" frequency.
In a similar way we obtain the space $\overline{\cal H}$ of negative
frequency from the solutions bounded in $M_R^+$. The spaces $\cal H$ and
$\overline{\cal H}$ are a direct decomposition of the vector space $V$ of
solutions in $M_L$. So any solution $\Phi$ in $M_L$ can be decomposed in a
positive and negative component (\mbox{ \( \Phi= \Phi^+ + \Phi^- \)} ) .
This gives the possibility to define a complex structure $J$ with \mbox{
\(J \Phi=i \Phi^+ +(-i) \Phi^- \)} on $\Sigma$, which is compatible with
the natural symplectic form

\begin{equation}
\Omega (\Phi,\Phi') = \int_{\Sigma} (  \Phi \partial_{\mu} \Phi'
                              - \Phi' \partial_{\mu} \Phi ) d \Sigma^{\mu}_L
\label{eq:spl2}
\end{equation}
on $M_L$.

To make $\cal H$ to a one-particle Hilbert space it remains to construct a
scalar product. Following \rf{eq:scalar-hilbert} we define
\begin{equation}
   \langle  \Phi_1, \Phi_2    \rangle  =
   i  \int_\Sigma
   (     \overline{\Phi_2} \partial_t \Phi_1
      -  \Phi_1 \partial_t \overline{\Phi_2} )
   d \Sigma_L .
\label{eq:scalar-tunnel}
\end{equation}
It remains to show, that this product is positive definite for
functions in $\cal H$. Therefore we consider $\Phi$ on $\Sigma$ and its
derivatives as the boundary values of a function $\Phi_R$ in $H^+$ and
$\overline{\Phi}$ and its derivative as boundary values of a function
$\Phi_{R \theta}$ in $H^-$. Thus on $\Sigma$ we have for the modes

\begin{equation}
   \frac{\partial \Phi}{\partial t} = -i
   \frac{\partial \Phi_R}{\partial \tau}
   \mbox{ \ \ \ \ \ \ and \ \ \ \ \ }
   \frac{\partial \overline{\Phi}}{\partial t} =  -i
   \frac{\partial \Phi_{R \theta}}{\partial \tau}.
\label{eq:derivative-on-sigma}
\end{equation}
If we choose the coordinate $x^0$ in $M_L^+$ and $M_R^-$ to be orthogonal
to $\Sigma$ it will be easy to see that the measures of integration, which
are the dual forms with respect to $M_L^+$ or $M_R^-$ of the volume 3-form
on $\Sigma$, are equal. We get

\begin{equation}
  d \Sigma_L = (-g_L^{00}) \sqrt{-g_L} d^3\!x
  = g_R^{00} \sqrt{g_R} d^3\!x = d\Sigma_R.
\label{eq:measure}
\end{equation}
This gives
\begin{equation}
   \langle  \Phi, \Phi  \rangle =
   \int_{\Sigma} (   \Phi_{R \theta} \, \partial_\tau \!\Phi_R -
      \Phi_R \, \partial_\tau \! \Phi_{R \theta} ) \, d\Sigma_R.
\label{eq:positive1}
\end{equation}
{}From Gauss' formula the integration over $\Sigma$ can be replaced by an
integration over $M_R^-$:
\begin{equation}
   \langle  \Phi, \Phi  \rangle =
   \int_{M_R^-} ( \Phi_{R \theta} \, \nabla^2 \!\Phi_R -
      \Phi_R \, \nabla^2 \! \Phi_{R \theta}   ) \, dV_R.
\label{eq:positive2}
\end{equation}
The function $\Phi_R$ is not bounded in $M_R^+$. It satisfies the
modified Klein-Gordon equation

\begin{equation}
         (-\nabla^2_{g_R} + m^2)\, \Phi_R =j.
\label{eq:klein-j}
\end{equation}
The source $j$ has support only in $M_R^+$. Similarly the function
$\Phi_{R \theta} $ has a source $j_\theta$ in
$M_R^-$ and we can write

\begin{equation}
    \langle    \Phi, \Phi  \rangle =
   \int_{M_R^-} \Phi_R j_\theta \, dV_R.
\label{eq:positive3}
\end{equation}
Let $G(x,y)$ be the unique inverse of the Klein-Gordon operator
\mbox{$   -\nabla^2_{g_R} + m^2  $ } on $2M_R$, then $\Phi_R(x)$ can be
written as:

\begin{equation}
   \Phi_R (x)  = \int_{M_R^+}G(x,y) \,j(y) \, dV_{Ry}
\label{eq:positive4}
\end{equation}
and we have
\begin{equation}
   \langle  \Phi, \Phi  \rangle =
   \int_{M_R^+} \int_{M_R^-}
   j(x) G(x,y) j(y)_\theta \, dV_{Rx} \, dV_{Ry}.
\label{eq:positive5}
\end{equation}
Because $j$ vanishes in $M_R^-$ and $j_\theta$ in $M_R^+$, we
may extend the integral to \mbox{ $ 2M_R \times 2M_R $ } and we can write
also:
\begin{equation}
   \langle  \Phi, \Phi  \rangle =
   \int_{2M_R \times 2M_R}
   j(x) G(x,\theta y) j(y) \, dV_{Rx} \, dV_{Ry}.
\label{eq:positive5a}
\end{equation}
We have a suitable scalar product \rf{eq:scalar-tunnel}, if the Green's
function $G(x,y)$ is such that the expression \rf{eq:positive5} is
non-negative for all possible sources $j$ e.g. those satisfying the
requirement of Reflection Positivity will give a satisfactory physical
inner product. The requirement of reflection positivity is quite
stringent. It is not sufficient for example that $G(x,y)$ be pointwise
positive. If one considers a Gaussian in flat one dimensional Euclidean space,

\begin{equation}
   G(x,y) = \exp \bigl( - \frac{(x-y)^2}{2} \bigr)
\label{eq:gauss}
\end{equation}
and take the source to be
\begin{equation}
   j(x)= x \exp \bigl( - \frac{(x-1)^2}{2} \bigr),
\label{eq:quelle}
\end{equation}
the integral turns out to be
\begin{equation}
   \int_{-\infty}^\infty \int_{-\infty}^\infty
   j(x) G(x,y) j(-y) d\!x d\!y =
   -\frac{4 \pi}{3^{5/2} e^{2/3}}.
\label{eq:positiv_integral}
\end{equation}
It can be seen not to satisfy positivity.

For the simple case of a free theory one may show directly that the
expression \rf{eq:positive5} is positive as follows.  We define the function
\begin{equation}
   G_D(x,y) =  G(x,y)   -  G(x,\theta y).
\label{eq:gd}
\end{equation}
While $G$ is the Green's function on $2M_R$, $G_D$ is the Green's function
on $2M_R$ restricted to functions which vanish on $\Sigma$. So as
operators it follows that \mbox{ $ G_D \leq G $ } and we can see that the
expression \rf{eq:positive5a} rewritten as
\begin{eqnarray}
   \langle  \Phi, \Phi  \rangle = & &
   \int_{2M_R \times 2M_R}
   j(x) G(x,y) j(y) \, dV_{Rx} \, dV_{Ry} \nonumber \\
  &-&
    \int_{2M_R \times 2M_R}
   j(x) G_D(x,y) j(y) \, dV_{Rx} \, dV_{Ry}
\label{eq:positive6}
\end{eqnarray}
is indeed positive, provided the $j$'s are in a class of functions
which guarantees the finiteness of the integrals.

Just as in Lorentzian space we can define suitable field-operators in the
Riemannian section $2M_R$. We set
\begin{equation}
   {\bf \Phi}_R   =  \sum_k
         (  {\bf a}_k \Phi_{R k}^- + {\bf a}^\dagger_k \Phi_{R k}^+ ),
\label{eq:field-operator-allg-riemann}
\end{equation}
where the normalized modes $\Phi_{R k}^\pm$ are elements of $H^\pm$.
The vacuum expectation value of the product of two field-operators gives
just the Riemannian Green's function:
\begin{equation}
   G(x,y)   =  \langle 0|{\bf \Phi}_R (x){\bf \Phi}_R (x)|0 \rangle  =
      \sum_k \Phi_{R k}^-(x) \Phi_{R k}^+(y)
\label{eq:schwinger}
\end{equation}
The connection between quantum field theory in Minkowski space and properties
of corresponding Green's functions in Euclidean space is well investigated.
Let us compare our results with Osterwalder and Schrader's approach to
Euclidean quantum field theory.
In their paper they gave necessary and sufficient conditions under which
Euclidean Green's functions have analytic continuations whose boundary
values define a unique set of Wightman distributions. These conditions
were

(E0) Temperedness

(E1) Euclidean covariance

(E2) Reflection positivity

(E3) Symmetry

(E4) Cluster property.

Caused by the curvature and possible finiteness of the Riemannian space
$2M_R$ in our approach we can not expect something analog to the
conditions of Euclidean covariance (E1) and the cluster property (E4), but
these conditions where anyway only included to guarantee similar
properties for the corresponding Wightman functions.  In our approach we
assumed that the integrals in \rf{eq:positive6} exist, which is in analogy
to (E0).  The way, we constructed the Euclidean Green's functions makes
them symmetric (E3) and guarantees the reflection positivity (E2). Then,
because of the fact, that the functions $ \Phi_R^\pm $ in $2M_R$ are a
complex continuation of the functions $ \Phi^\pm $, the corresponding
Wightman function \rf{eq:wightman} is a continuation of the Riemannian
Green's function. In this sense our paper presents a generalization of the
results of Osterwalder and Schrader in the case of tunneling geometries.

\section{Examples}
\setcounter{equation}{0}
\label{sec-example}
\subsection{DeSitter space}

The  simplest  case  of  a  tunneling  geometry  is  deSitter  space.  The
complexified space can be considered as the surface given by
\begin{equation}
     (z^{1})^{2}  + (z^{2})^{2}  + (z^{3})^{2}
   + (z_{4})^{2}  +(z_{5})^{2}  = 1 .
\label{eq:kugel}
\end{equation}
The Riemannian real part \(2M_{R}\) is given by the involution
\begin{equation}
   J_{R}: (z^{1},z^{2},z^{3},z^{4},z^{5})
   \rightarrow (\overline{z^{1}},
           \overline{z^{2}},
                     \overline{z^{3}},
                \overline{z^{4}},
                     \overline{z^{5}}),
\label{eq:involution_r}
\end{equation}
which is a four sphere \(S_{3}\). We may consider \(M_{R}^{-}\) as the
lower  and  \(M_{R}^{+}\)  as  the
upper half sphere. The Lorentzian real part \(2M_{L}\) is given by the
involution
\begin{equation}
        J_{R}: (z^{1},z^{2},z^{3},z^{4},z^{5})
   \rightarrow (\overline{z^{1}},
           \overline{z^{2}},
                     \overline{z^{3}},
                \overline{z^{4}},
                     -\overline{z^{5}}).
\label{eq:involution_l}
\end{equation}
The intersection \(\Sigma=2M_{R}\cap 2M_{L}\) of both spaces is a three
sphere.

   With suitable coordinates the metric of the Riemannian  section  can
be cast in the form
\begin{equation}
        ds^{2}=-\frac{1}{\cosh\!\!^{2}\tau}(d\tau^{2}+d\Omega_{3}^{2})
\label{eq:line_element_desitter}
\end{equation}
where \(d\Omega_{3}^{2}\) is the metric of the sphere \(S_{3}\) and
\(\tau\in(-\infty,\infty)\). The  boundary \(\Sigma\)
is given by \(\tau=0\). In  these  coordinates  the  Klein-Gordon  equation
\rf{eq:klein_gordon2} takes the form
\begin{equation}
        -\cosh\!\!^{4}\tau
        \frac{\partial}{\partial\tau}
   (\frac{1}{\cosh\!\!^{2}\tau}
   \frac{\partial \Phi_R}{\partial\tau})
   -\cosh\!\!^{2}\tau
   \Delta_{3} \Phi_R
   +m^{2}   \Phi_R=0
\label{eq:tagirov_equation}
\end{equation}
and we assume $m^{2}>9/4$.

   By a separation of variables we find the following set of linearly
independent solutions:
\begin{equation}
   \Phi_{R\,pqr}^{\pm}\!(\tau,\Omega;m)=
   y_{Rp}^{\pm}\!(\tau;m)\,{\cal Y}_{pqr}\!(\Omega)
\label{eq:modes}
\end{equation}
\centerline{\(  p = 0,1,2,\cdots \; ; \;\;
      q = 0,1,\cdots,p \; ; \;\;
      r = -q,\cdots,q  \;.
\)}
The index \(R\) indicates that we have a mode in Riemannian space and the
\( {\cal Y}_{pqr}\!(\Omega)   \)
are ortho-normal real-valued surface harmonics of degree \(p\) on \(S_{3}\)
 obeying
\begin{equation}
   -\Delta_{3}{\cal Y}_{pqr}\!(\Omega)=
   p(p+2)\,{\cal Y}_{pqr}\!(\Omega) \;.
\label{eq:harmonics}
\end{equation}
The real valued functions
\( y_{Rp}^{\pm}\!(\tau;m)  \) may be expressed  in terms of a
hypergeometric function
\begin{eqnarray}
   y_{R\,p-1}^{\pm}\!(\tau;m)& = &
   \frac{1}{p!}
   \bigl( \,\Gamma(p+1/2+i\gamma)
    \,\Gamma(p+1/2-i\gamma)\, \bigr)^{1/2}
   \!\cosh \!\tau
   \exp(\pm p \tau)  \nonumber \\
   & & _{2}\! F_{1}
   \bigl( \, 1/2+i\gamma,\,1/2-i\gamma,\,p+1;
   \,\exp(\pm\tau)/(2\cosh\!\tau)\, \bigr)
\label{eq:tagirov_solution}
\end{eqnarray}
where
\(    \gamma = (m^{2} - 9/4)^{1/2}   \)
\cite{tagirov}. According to our  definition  the
\(\Phi_{R\,pqr}^{-}\!(\tau;m)\)
describe positive frequency modes, which are bounded for  \( \tau<0 \).
The  relation
\mbox{   \( \theta \, \Phi_{R\,pqr}^{+}~\!=~\!\Phi_{R\,pqr}^{-}   \)    }
holds true.

   By rotating back the time axis we get the positive frequency modes
\begin{equation}
   \Phi_{L\,pqr}^{+}\!(t;m)=\Phi_{R\,pqr}^{+}\!(-i\tau;m)
\label{eq:Lorentz_modes}
\end{equation}
and their span gives the set of positive frequency functions in  the
Lorentzian section of deSitter space. With this unique definition of positive
frequency we can expand the quantized Klein-Gordon field \(\Phi\) in \(2M_{L}\)
in a Fock space representation
\begin{equation}
   {\bf\Phi}_{L}(t,\Omega ;m) = \sum_{pqr}
   ({\bf a}_{pqr}\Phi_{L\,pqr}^{-}(m) +
   {\bf a}_{pqr}^{\dagger}\Phi_{L\,pqr}^{+}(m) \,)
\label{eq:field_operator_l}
\end{equation}
and the vacuum state would then be uniquely specified by
\( {\bf a}_{pqr}|0 \rangle\,=0  \).
We shall now elaborate this example further because we think it contains many
features which can be generalized.
%%%%%%%

In analogy to formula \rf{eq:field_operator_l}
we define in Riemannian space:
\begin{equation}
   {\bf \Phi}_{R}(\tau,\Omega ;m) = \sum_{pqr}
   ({\bf a}_{pqr}\Phi_{R\,pqr}^{-}(m) +
   {\bf a}_{pqr}^{\dagger}\Phi_{R\,pqr}^{+}(m) \,)\, .
\label{eq:field_operator_r}
\end{equation}
In equation \rf{eq:field_operator_r} the operators
      \( {\bf a}_{pqr}  \)
and
      \( {\bf a}_{pqr}^{\dagger} \)
are the same as those in \rf{eq:field_operator_l} and they act on the same
Hilbert space, which consists of suitable boundary data on $\Sigma$.

A general one-particle state is given by the supervposition
\begin{equation}
   |1 \rangle =\sum_{pqr} \alpha_{pqr}{\bf a}_{pqr}^{\dagger}|0\rangle.
\label{eq:superposition}
\end{equation}
This can also be expressed in terms of the field operators in Lorentzian space
as
\begin{eqnarray}
   |1 \rangle  &=&  \int_{M_L^+} dV_L
                  \alpha_L\!(t,\Omega) {\bf \Phi}_L\!(t,\Omega;m)|0\rangle
                                                                  \nonumber \\
               &=&  \sum_{pqr} \int_{M_L^+} dV_L \alpha_L\!(t,\Omega)
                  \Phi_{Lpqr}^+(t,\Omega;m) {\bf a}_{pqr}^{\dagger}|0\rangle,
\label{eq:super_lorentz}
\end{eqnarray}
and in Riemannian space
\begin{eqnarray}
   |1 \rangle  &=&  \int_{M_R^-} dV_R
                {\bf \Phi}_R\!(\tau,\Omega;m)|0\rangle
                                                                  \nonumber \\
               &=&  \sum_{pqr} \int_{M_R^-} dV_R \alpha_R\!(\tau,\Omega)
               \Phi_{Rpqr}^+(\tau,\Omega;m) {\bf a}_{pqr}^{\dagger}|0\rangle.
\label{eq:super_riemann}
\end{eqnarray}
What is the relation between the functions $\alpha_L\!(t,\Omega)$ and
$\alpha_R\!(\tau,\Omega)$?
Caused by the orthogonality of the Lorentzian modes of different mass
\cite{tagirov}
\begin{eqnarray}
   \int_{M_{L}^+}\!dV_{L}
   \,\Phi_{L\,pqr}^{-}(t,\Omega;m)
   \,\Phi_{L\,p'q'r'}^{+}(t,\Omega;m')=&
   \delta_{pqr,p'q'r'}/2   \nonumber   \\
   &\times \left\{ \begin{array}{ll}
      \delta(m^{2}-m'^{2})&\mbox{if $m,m'>3/2$} \\
      \infty          &\mbox{if $m,m'<3/2$}  \nonumber
   \end{array}
   \right.
\label{eq:orthogonality}
\end{eqnarray}
we can rewrite the Riemannian expression \rf{eq:super_riemann} as
\begin{eqnarray}
     |1 \rangle = && \sum_{pqr} \int_{M_R^-} dV_R \alpha_R\!(\tau,\Omega)
                     \sum_{p'q'r'} \int_{m'^2>9/4} dm'^2 \int_{M_L^+} dV'_L
                                                                \nonumber \\
                  && \Phi_{Rp'q'r'}^+(\tau,\Omega;m')
          \Phi_{Lp'q'r'}^-(t',\Omega';m') \Phi_{Lpqr}^+(t',\Omega';m)
                     {\bf a}_{pqr}^{\dagger}|0\rangle.
\label{eq:delta_einfugen}
\end{eqnarray}
By indroducing the Green's function
\begin{eqnarray}
   G(\tau'\!, \,\Omega'\!; \,\tau\!, \,\Omega\!; \,m')&=&
   \langle 0| {\bf \Phi}_{R}(\tau', \,\Omega'; \,m')
   {\bf \Phi}_{R}(\tau, \,\Omega; \,m')
   |0 \rangle    \nonumber   \\
   &=&  \sum_{p'q'r'}
   \Phi_{R\,p'q'r'}^{-}\!(\tau',\Omega';m')\,
   \Phi_{R\,p'q'r'}^{+}\!(\tau,\Omega;m')
\label{eq:Schwinger}
\end{eqnarray}
and analytic continuation we can rewrite our one-particle state as
\begin{eqnarray}
     |1 \rangle = && \sum_{pqr} \int_{M_L^+} dV_L \int_{m'^2>9/4} dm'^2
                     \int_{M_R^-} dV'_R
                                                                \nonumber \\
                  && G(-it,\Omega;\tau',\Omega';m')  \alpha_R\!(\tau',\Omega')
                     \Phi_{Lpqr}^+(t,\Omega;m)
                     {\bf a}_{pqr}^{\dagger}|0\rangle.
\label{eq:one_green}
\end{eqnarray}
A comparison with \rf{eq:super_lorentz} showes that we can express the
Lorentzian function $\alpha_L$ in terms of the Riemannian function
$\alpha_R$ by means of
\begin{equation}
   \alpha_L(t,\Omega)=  \int_{m'^2>9/4} dm'^2
                        \int_{M_R^-} dV'_R
                        G(-it,\Omega;\tau',\Omega';m')\alpha_R(\tau',\Omega').
\label{eq:alpha_l}
\end{equation}
We see that we can construct a one-particle state from the field operators
${\bf \Phi}_R$ defined in the Riemannian sector as well as from the field
operators in the Lorentzian sector.

\subsection{Page Metric}
The Page metric is a solution of the Einstein equation with the  cosmological
constant $\Lambda$. It belongs to the Bianchi IX  type  solutions  and  is
invariant under a group homomorphic to
\(U(1) \times SU(2)\). The line element  can  be  expressed
in terms of the coordinates
\( \eta,\psi,\theta,\phi   \) as \cite{gibbons_pope}
\begin{equation}
   ds^{2}=a^{2}b^{2}c^{2} d\eta^{2}
   + a^{2} d\sigma_{1}^{2}
   + b^{2} d\sigma_{2}^{2}
   + c^{2} d\sigma_{3}^{2}
\label{eq:line_element_page}
\end{equation}
Here the
\( \sigma_{1},\sigma_{2},\sigma_{2}  \)
are left-invariant one forms on $   SU(2) $ such that
\begin{eqnarray}
   \sigma_{1}&=&-\sin \! \psi \,d\theta +
      \cos\! \psi \sin\! \theta \,d\phi   \nonumber   \\
   \sigma_{2}&=&\cos\! \psi \,d\theta +
      \sin\! \psi \sin\! \theta \,d\phi   \nonumber   \\
   \sigma_{3}&=& d\psi + \cos\! \theta\, d\phi  \nonumber
\label{eq:left_forms}
\end{eqnarray}
The functions $a,b,c$ are given by the relations \cite{page}
\begin{eqnarray}
   a^{2}=b^{2}&=&\frac{a_{0}^{2}}{\lambda}
   \,(1-\nu^{2}\tau^{2})      \label{eq:abc1}      \\
   c^{2}&=&\frac{c_{0}^{2}}{\lambda}
   \,\frac{\Delta(\tau)}{(1-\nu^{2}\tau^{2})}
   \,(1-\tau^{2})       \label{eq:abc2}   \\
   d\eta&=&\frac{\lambda}{a_{0}^{2}c_{0}\Delta(\tau)
   \,(1-\tau^{2})}\,d\tau     \label{eq:abc3}
\end{eqnarray}
with
\( (-1 < \tau < 1)    \).
The constant \(\nu\) is the solution \(\nu\approx0.28\ldots\) of the equation
\begin{equation}
     \nu^{4} + 4\nu^{3} - 6\nu^{2} + 12\nu - 3 = 0
\label{eq:zero1}
\end{equation}
and we used the definitions
\begin{eqnarray}
   \Delta(\tau)&=& 3 - \nu^{2} -
   \nu^{2}(1+\nu^{2})\tau^{2}    \nonumber   \\
   \lambda&=& \frac{\Lambda}{3(1+\nu^{2})}
\label{eq:delta_lambda}
\end{eqnarray}
and
\begin{eqnarray}
   a_{0}^{2}&=&\frac{1}{3+6\nu^{2}-\nu^{4}}  \nonumber   \\
   c_{0}^{2}&=&\frac{1}{(3+\nu^{2})^{2}}  \nonumber
\label{eq:a0_c0}
\end{eqnarray}
For real $\tau$ the equations
\rf{eq:abc1}, \rf{eq:abc2} and \rf{eq:abc3}
give the compact Riemannian section
$2M_{R}$. The reflexion map is given by
\( \theta:\tau \rightarrow -\tau \) and thus the surface
\( \tau=0   \) is  our
nucleation surface $\Sigma$, which is a squashed three-sphere. To obtain
the Lorentzian section we have to take $\tau$ to be pure imaginary.
We get an ever expanding
universe in which $a$ and $c$ grow exponentially as
\( \exp(\Lambda\tau/3)  \).

With the help of the vector fields \cite{hu}
\begin{eqnarray*}
   \xi_{1}&=&- \cot\!\theta \cos\!\psi \frac{\partial}{\partial\psi}
   - \sin\!\psi \frac{\partial}{\partial\theta}
   + \frac{\cos\!\psi}{\sin\!\theta} \frac{\partial}{\partial\phi}   \\
   \xi_{2}&=&- \cot\!\theta \sin\!\psi \frac{\partial}{\partial\psi}
   + \cos\!\psi \frac{\partial}{\partial\theta}
   + \frac{\sin\!\psi}{\cos\!\theta} \frac{\partial}{\partial\phi} \\
   \xi_{3}&=&\frac{\partial}{\partial\psi}
\label{eq:xi_s}
\end{eqnarray*}
the Klein-Gordon equation can be written as
\begin{equation}
   -\frac{1}{a^{2}b^{2}c^{2}}\frac{\partial^{2}}{\partial\eta^{2}} \Phi
   -( \frac{\xi^{2}_{1}}{a^{2}}+
      \frac{\xi^{2}_{2}}{b^{2}}+
      \frac{\xi^{2}_{3}}{c^{2}}  )\,\Phi
   +m^{2}\Phi=0
\label{eq:klein}
\end{equation}
A separation of variables is possible and we can use the following ansatz for
the linearly independent solutions
\begin{equation}
   \Phi^{(i)}_{R\,pqr}(\tau,\psi,\theta,\phi)=
   y_{R\,pq}(\tau)\,D_{qr}^{(i)p}\!(\psi,\theta,\phi)\;,
\label{eq:ansatz}
\end{equation}
where $i=1,2$.  The functions
\( D_{qr}^{(i)p}  \) are the real and the imaginary part of the
Wigner functions \( D_{qr}^{p}=D_{qr}^{(1)p}+iD_{qr}^{(2)p} \)
\cite{dawydov}. They satisfy the relations
\begin{equation}
   \xi_{3}D_{qr}^{p}=iqD_{qr}^{p}\;,
\label{eq:xi3}
\end{equation}
\begin{equation}
        (\xi_{1}^{2}  + \xi_{2}^{2} + \xi_{3}^{2} )\,
        D_{qr}^{p}=-p(p+1)D_{qr}^{p}
\label{eq:xi^2}
\end{equation}
and
\begin{equation}
   \sum_{p} \overline{D_{qr}^{p}}D_{qr'}^{p}
   =\sum_{p} \overline{D_{rq}^{p}}D_{r'q}^{p}=
   \delta_{rr'}
\label{eq:wigner_orthogonal}
\end{equation}
The ansatz \rf{eq:ansatz} together with the relations
\rf{eq:xi3} and
\rf{eq:xi^2} leads to the equation
\begin{equation}
   -\frac{1}{a^{4}c^{2}}
   \frac{\partial^{2}} { \partial\eta^{2} }\,    y_{R\,pq}
   + (\, \frac{p(p+1)}{a^{2}}
      + q^2 ( \frac{1}{c^{2}} - \frac{1}{a^{2}} \, )\;
   ) \,y_{R\,pq} + m^{2}y_{R\,pq}=0
\label{eq:page_equation}
\end{equation}
With the substitution $\tau^{2}=z$ we get
\begin{eqnarray}
   &&-y''_{R\,pq} -
   ( \frac{1}{2z} + \frac{\Delta'}{\Delta} -\frac{1}{1-z} )\,y'_{R\,pq}
   + \frac{1-\nu^{2}z}{4z(1-z)\Delta}     \\
   && \times (\frac{p(p+1)}{a_{0}^{2}(1-\nu^{2}z)}
   + q^{2}  \,( \frac{1-\nu^{2}z}{c_{0}^{2}(1-z)\Delta}
    -\frac{1}{a_{0}^{2}(1-\nu^{2}z)}
   ) + \frac{m^{2}}{\lambda}     ) \,y_{R\,pq} = 0 \nonumber
\label{eq:page_equation_z}
\end{eqnarray}

The 2-surface "bolt" $z=1$ which closes up the space corresponds to a regular
singular point of the differential equation \rf{eq:page_equation_z}. With the
ansatz \cite{ince}
\begin{equation}
   y_{R\,pqr}=(1-z)^{k}u(z)
\label{eq:entwicklung_z=1}
\end{equation}
where $u$ is to be considered analytic and different from zero at $ z=1$,
we  get the indicial equation
\begin{equation}
   k^{2}-2k-\frac{(1-\nu^{2})^{2}}{4c_{0}^{2}\Delta^{2}(1)}q^{2}=0
\label{eq:indicial_1}
\end{equation}
which has the solution
\begin{equation}
      k^{\pm}  = 1 \pm
   \sqrt{   1 +
      \frac { (1-\nu^{2})^{2} }{ 4c_{0}^{2}\Delta^{2}(1)} q^{2}  }
   \approx 1 \pm \sqrt{ 1 + 3.07 q^{2}}  .
\label{eq:indicial_solution}
\end{equation}
We see that there is at least one solution which is regular at  $z=1$.
Because equation
\rf{eq:page_equation_z}
has no other singularity in the circle $|1-z|<0$ and our space $2M_{R}$
is covered by the interval $0<z<1$  it  remains  to  consider  the
point  $z=0$.
According to Fuchs's theory this point also represents  a  regular  singular
point of the differential equation and we use an ansatz  similar  to
\rf{eq:entwicklung_z=1},  in
which we assume that the function $v(z)$ is analytic and different from zero
at $z=0$
\begin{equation}
   y_{R\,pqr}=(z)^{l}v(z)
\label{eq:entwicklung_z=0}
\end{equation}
and we get
\begin{equation}
   l(l-\frac{1}{2})=0
\label{eq:indicial_equation_l}
\end{equation}
{}From the properties of the function $v$ we see, that there is no singularity
in the general solution of equation
\rf{eq:page_equation_z} at $z=0$.  Together  with  what  we  found
about the behaviour of the solutions around $z=1$ and transforming back to
the parameter $\tau$ we conclude that there exist of solution
$y_{R\,pqr}$
which are everywhere regular in the interval $0>\tau>-1$. These can be used to
construct the positive frequency modes via the ansatz
\rf{eq:ansatz}.

\section{Spinors}
\label{sec-spinor}
\setcounter{equation}{0}

In this section we shall treat the spinorial case with what may seem to the
reader to be positively painful pedantry. However in view of the selection
rule found in \cite{gibbons-hawking} in the purely Lorentzian theory we feel
that this is justified since there appears to be a genuine difference between
the Lorentzian and the Riemannian theory (For example in the Lorentzian
theory there is no Lorentz -cobordism of the $S^3$ admitting an $SL(2,C)$
spinor structure. On the other hand one can clearly consider $S^3$ as the
boundary of the 4-ball , $B^4$ and put a Riemmannian metric on it). In
calculations of the wave function of the universe there appears to be no
obstacle to including spinors and indeed this has been done by Halliwell and
D'Eath~\cite{eath}. This contrast between the Riemannian and the Lorentzian
theory is rather puzzling.  We shall in fact find, in accordance with the
results of Halliwell and D'Eath that there is no apparent difficulty here.

The treatment sketched below is related to that of \cite{osterwalder-spinor},
\cite{osterwalder-frohlich} and \cite{nicolai} but we mentioned in
section~\ref{sec-complex} our strategy is to start with Majorana spinors and
then pass to Dirac spinors just as for spin zero particles we started with a
real scalar field and then passed to the charged case. Thus we shall use
conventions for the $\gamma $ -matrices $\gamma ^ {\mu}$ in Minkowski
space-time as follows:
\begin{equation}
   (\gamma_L ^0)^2   =  -1,   \ \ \ \ \
   (\gamma _L ^i )^2 =  1.
\label{eq:spin1}
\end{equation}
The matrices $\gamma^ \mu $ may be taken to be real with $\gamma ^0$
anti-symmetric and $\gamma ^i$ symmetric. The Dirac equation is thus
\begin{equation}
   (\gamma ^{\mu }_L D_ {\mu } - m) \psi =0.
\label{eq:spin2}
\end{equation}
where $\psi$ is a real
four-component spinor. Given a Cauchy surface $\Sigma$ the restriction of
$\psi$ to $\Sigma$ gives four real functions which constitute the real vector
space $V$ of cauchy data for the Dirac equation. The real vector space
$V$ admits an invariant positive definite inner product:
\begin{equation}
   \int _ \Sigma {\overline \psi} \gamma _L ^{\mu }\psi d \Sigma _{\mu} =
   \int _ \Sigma \psi ^t \psi  \sqrt h d ^ 3 x ,
\label{eq:spin3}
\end{equation}
where $\overline \psi$ is the Dirac adjoint, which coincides with the
Majorana adjoint in our case. The problem of quantization in this case is to
endow $V$ with a complex structure (or equivalently a symplectic
structure) compatible with the positive definite inner product. As before we
complexify $V$ and consider complex-valued (i.e. Dirac) spinors and
seek an orthogonal direct sum decomposition
\begin{equation}
   V_{C} = H_1 \oplus {\overline H_1}.
\label{eq:spin4}
\end{equation}
The Cauchy data may also be regarded as complex valued boundary data for the
Riemannian Dirac equation:
\begin{equation}
   (\gamma ^{\mu}_R D_{\mu} -m) \psi =0,
\label{eq:spin5}
\end{equation}
now,
\begin{equation}
   (\gamma _R ^0)^2 =1.
\label{eq:spin6}
\end{equation}
In fact choosing:
\begin{equation}
   \gamma_R^0 =i \gamma_L^0
\label{eq:spin7}
\end{equation}
one has that the $\gamma_R^{\mu}$ are hermitean. The Dirac adjoint in
$M_R$ may thus be taken as Hermitean conjugation. The conserve d inner
product, which arises from the conserved current on $M_R$ is:
\begin{equation}
   \int _{\Sigma} \psi ^{\dagger} \gamma^{\mu}_R \psi d \Sigma _{\mu}= \int
_{\Sigma} \psi ^ {\dagger}
i \gamma^0_R \sqrt h d^3x
\label{eq:spin8}
\end{equation}
It is important to realize that this expression
is {\sl not } the same as the analytic continuation of the Lorentzian inner
product.
Thus it cannot be regarded as the quantum-mechanical metric. This will be
defined below.One can now decompose $V_C $ into data $\in H^+$ which
is non-singular everywhere on $M_R^-$ and that $\in H^+$
which is everywhere non-singular on $M_R^+$. We can extend $\theta$ to a map
\begin{equation}
   \theta:  H^{\mp}
\rightarrow H^{\pm}
\label{eq:spin9}
\end{equation}
by defining;
\begin{equation}
   \psi_ {\theta} =\gamma ^0_R \psi (\theta x)
\label{eq:spin10}
\end{equation}
and the inner product to be:
\begin{equation}
   \int_{\Sigma} \psi _{\theta} ^{\dagger} \gamma^0_R \psi \sqrt h d^3 x.
\label{eq:spin11}
\end{equation}
Clearly this expression (i.e. \rf{eq:spin11} ) unlike that in
\rf{eq:spin8} coincides with the usual
Lorentzian inner product when restricted to the appropriate set of Cauchy
data.

Let us now turn to the case of Dirac spinors. It is here that the notation
can become confusing and the idea of doubling enters. We must start with
classical solutions of the Dirac equation. These are already "complex valued".
The associated complex conjugation operator is of course what one calls
"charge conjugation".  (In the Majorana
representation that we are using there is no need for an explicit charge
conjugation matrix). However this charge conjugation operator when extended
to the quantum theory is expected be a linear rather than an antilinear
operator. It follows that strictly speaking that Dirac spinors do not take
their values in the usual complex numbers of quantum mechanics but in a
commutative field (in the algebraic sense of those words) which is of course
isomorphic but not naturally so in the mathematical sense and not physically
equivalent. We could adopt the notation suggested earlier and introduce a new
imaginary unit $ i ^ {\prime}$.  If the Dirac spinors corresponded to
electrons in QED for example the new unit $ i^{\prime} $ corresponds to an
electromagnetic gauge transformation of $\pi \over 2$. Now let us turn to
the quantum theory. We must further complexify the space of classical
solutions of the Dirac equation by taking the tensor product with the usual
quantum mechanical complex numbers. We then impose on this extended space a
choice of positive and negative frequency, i.e. a complex structure, in the
same way that did for Majorana spinors using the reflection map $\theta$.
Using the notation $i$ for this complex structure we have that $i$ and
$i^\prime$ commute.

The problem becomes more confusing when we want to consider Weyl spinors. As
pointed out earlier, in Lorentzian spacetime, pointwise, we may obtain Weyl
spinors by using $\gamma _5$ as a complex structure. This means that we need
to act on the Dirac spinors with the projection operator:
$$
{1 \over 2}
(1-i ^ {\prime }\gamma _5)
$$
to obtain Weyl spinors satisfying:
$$
\gamma _ 5 \psi = i ^ {\prime } \psi.
$$
Now since $\gamma _5$ does not commute with $\gamma _0$ ,the two complex
structures, corresponding to the distinction between two different
chiralities and between particle and antiparticle respectively do not
commute.

\section{Conclusion}
\label{sec-conclusion}
\setcounter{equation}{0}
In section~\ref{sec-complex} we summarized the relation between the direct
sum decomposition of the space of real classical solutions of the
Klein-Gordon system with the choice of a complex structure. Then we
focused our attention to the question how the complex structure might be
related with geometrical properties of spacetime. If we follow the ideas
of the no-boundary proposal and the "tunneling of the universe from
nothing" then we have to assume a close physical relation between
Lorentzian and Riemannian sections of space. So the analytic continuation
of a Lorentzian spacetime to a Riemannian section is not only formal and
we followed the ideas of Euclidean field theory. We found that we could
relate analytic properties of the classical solutions in a Riemannian
section of space with a preferred complex structure on the nucleation
surface. So in a more popular way we can say, that in our construction
with the beginning of time a preferred decomposition of matter in
particles and anti-particles is given, which is strictly related with
non-local properties of wave-mechanics "before the creation of the
universe".

\end{document}